# Nonstandard electron dynamics in topological insulators subjected to magnetic field: the Berry phase effects

*V.Ya. Demikhovskii* [1], *R.V. Turkevich*

University of Nizhny Novgorod, 603950 Nizhny Novgorod, Russian Federation
[1] demi@phys.unn.ru

The quasiclassical dynamics is studied for charge carriers moving on the surface of 3D topological insulator of $Bi_2Te_3$ type and subjected to static magnetic field. The effects connected to the symmetry changes of electron isoenergetic surfaces (contours) and to the nonzero Berry curvature are taken into account. It is shown that in contrast to the standard dynamics of the electrons moving in constant and uniform magnetic field along the trajectories defined by the equations $E(k)$=const and $p_z$=const, here some new effects are arising, being related to both the appearance of the anomalous velocity term proportional to the Berry curvature, and to the trajectory bending related to the additional term for the energy proportional to the orbital momentum of the wavepacket. This should lead to the changes in cyclotron resonance conditions of the surface electrons. Although the time reversal invariance and the topological order are broken in the magnetic field, the investigation of cyclotron resonance allows determining whether this insulator was trivial or nontrivial at zero magnetic field.

As it follows from the theoretical and experimental studies performed during the last years, the Berry curvature and the orbital magnetic moment of the wavepackets along with the band spectrum of carriers define their dynamics in the external electromagnetic field, and also in the deformed crystal [1], [2]. Here, as it was shown in papers [3], [4], the temporal evolution of the mean coordinate $r_c(t)$ and mean momentum $p_c(t)$ of the wavepacket is determined from the quasiclassical equations of motion. Including the Berry curvature is also necessary during the calculations of many other important characteristics and, in particular, the nontrivial conditions of cyclotron orbit quantization in the magnetic field [5]. In the present paper we consider the equations of motion for the wavepacket in 3D topological insulator of $Bi_2Te_3$ type subjected to external static and uniform magnetic field.

As it is known, the energy spectrum of charge carriers in topological insulators is characterized by several peculiarities. For example, the isoenergetic contours for the electrons on the surface of the topological insulator transform from the circles into the sixfold symmetry curves under the influence of the periodic potential of crystal lattice. Such effect has been discussed by Fu [6]. The hexagonal warping of the Fermi surface in $Bi_2Te_3$ [7], $Bi_2Se_3$ [8] and $Pb(Bi,Se)_2Te_4$ [9] topological insulators has been discovered by using the ARPES technique, i.e. the angle resolved spectroscopy.

In the present paper the effects of nonstandard electron dynamics are considered for the wavepackets composed from the surface states in topological insulators, without including the particle spin. The electron states in the $Bi_2Te_3$ with inclusion of the hexagonal warping of the electron spectrum of surface states, according to [6], are described by the Hamiltonian

$$H = \hbar v_f\big(k_x \sigma_y - k_y \sigma_x\big) + \hbar \frac{\lambda}{2}(k_+^3 + k_-^3)\sigma_z + \Delta \sigma_z \qquad (1)$$

which contains the term, proportional to the third power of *k*. The parameters are $\lambda = 3.7 \cdot 10^7$ cm$^3$/s, $v_f = 3.86 \cdot 10^7$ cm/s, $k_\pm = k_x \pm i k_y$, and $\sigma_i$ are the Pauli matrices. This Hamiltonian has the $C_{3V}$ symmetry. The electron spectrum of Hamiltonian (1)

$$E_0(k_x, k_y) = \pm\sqrt{\hbar^2 k^2 v^2 + (\lambda k^3 Cos(3\theta) + \Delta)^2}. \qquad (2)$$

Here $\theta$ is the polar angle, and the + and − signs correspond to the conduction band and the valence band, respectively. In this model the electron spin is not taken into account.

The Lagrangian of a moving wavepacket centered at $\vec{r}_c$ и $\vec{k}_c$ can be written in the form

$$L(\vec{r}_c, \vec{k}_c, t) = \left\langle \Psi \left| i\hbar \frac{d}{dt} - \hat{H} \right| \Psi \right\rangle, \tag{3}$$

where $|\Psi\rangle$ is the wavepacket state. The Euler-Lagrange quasiclassical equations for $\vec{r}_c$ и $\vec{k}_c$ describing the motion of the two-dimensional packet along the cyclotron orbit were obtained in [3,4] by the minimization of the action functional

$$S[\vec{r}_c(t), \vec{k}_c(t)] = \int dt' L(t'). \tag{4}$$

These equations have the following form:

$$\dot{\vec{k}} = -\frac{e}{c\hbar}[\dot{\vec{r}} \times \vec{B}], \tag{5a}$$

$$\dot{\vec{r}} = \frac{1}{\hbar}\frac{\partial E_m}{\partial \vec{k}} - [\dot{\vec{k}} \times \vec{\Omega}], \tag{5b}$$

where $\vec{B}$ is the vector of the external magnetic field, $\vec{\Omega} = (0,0,\Omega_z)$ is the Berry curvature which has only the z-component in our case. For the standard equations of motion in the magnetic field the electron trajectory is defined from the conditions $E_0$=const and the constant projection of the momentum on the magnetic field direction $p_z$=const. In contrast to the standard case for the model with Hamiltonian (1) here the energy $E_m = E_0 - \vec{m}\cdot\vec{B}$ is conserved, and the trajectory is defined from the condition $E_m$=const. The energy $E_m$ consists from two terms: $E_0$ is the unperturbed electron energy, and $-\vec{m}\cdot\vec{B}$ is the energy of the wavepacket rotating in the magnetic field around its mass center. Besides, as it follows from the equation (5b), the electron velocity acquires the additional term proportional to the z-component of the Berry curvature.

In topological insulators where the electrons are described by the Hamiltonian (1) the explicit expressions for the Berry curvature and orbital magnetic moment have the following form:

$$\Omega_z(k_x, k_y) = -\frac{\hbar^2 v^2(\Delta - 2\lambda\hbar k_x^3 + 6\lambda\hbar k_x k_y^2)}{2(\hbar^2 v^2(k_x^2 + k_y^2) - (\Delta + \hbar\lambda(k_x^3 - 3k_x k_y^2))^2)^{\frac{3}{2}}}, \tag{6}$$

$$m_z(k_x, k_y) = -\frac{e}{2\hbar c}\frac{\hbar^2 v^2(\Delta - 2\lambda\hbar k_x^3 + 6\lambda\hbar k_x k_y^2)}{(\hbar^2 v^2(k_x^2 + k_y^2) - (\Delta + \hbar\lambda(k_x^3 - 3k_x k_y^2))^2)}. \tag{7}$$

The Berry curvature and the magnetic moment in the valence band and in the conduction band are related to each other as $\Omega_V(k) = -\Omega_C(k)$, $m_V(k) = m_C(k)$.

Excluding the dimensional constant, the expression (6) for the Berry curvature differs from expression (7) for $m_z$ only by the value of power for the expression in parenthesis in the denominator (the power is 3/2 instead of 1). If one excludes the effect of trajectory bending in the Hamiltonian, than the Berry curvature and the orbital magnetic moment are both proportional to the parameter $\Delta$. If the trajectory bending is taken into account, the Berry curvature and the orbital magnetic moment do not vanish even for $\Delta$=0. The Berry curvature will change its sign six times while passing along the cyclotron trajectory in **k**-space. From (2) and (7) it follows that if the unbiased spectrum $E_0$ has the sixfold rotational axis, the term $-\vec{m}\cdot\vec{B}$ will have the threefold symmetry axis. The magnetic energy is typically lower than the energy of the electron spectrum and can be neglected, however in strong magnetic field (B ~ 1 MGs for $Bi_2Te_3$) this energy correction becomes significant.

It should be noted that while the standard equations of charged particle in the magnetic field do not contain the Plank constant, the equations (5a), (5b) include additional terms related to the virtual transitions into neighboring bands, namely, the anomalous velocity term $-\left[\dot{\vec{k}} \times \vec{\Omega}\right]$ and the additional electron energy in the magnetic field $-\vec{m}\cdot\vec{B}$, both are proportional to the first power of the Plank constant $\hbar$.

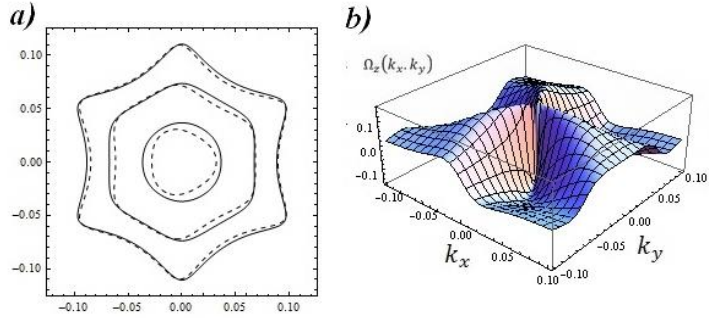

Fig.1. (a) – Isoenergetic contours $E_0$=const (bold lines) and $E_m$=const (dashed lines). (b) – Berry curvature measured in units of $\Omega_z/l_b^2$ (at $B$=1 MGs), $k$ is in units of $Å^{-1}$.

In this case the energy $E_m$ will have the threefold rotational axis, and the isoenergetic lines will have the shape which is shown in Fig.1(a). The second difference of the system of equations (5a), (5b) from the conventional equations is the presence of the additional velocity term in the second equation, which is proportional to the Berry curvature in Fig.1(b). It should be noted that the Berry curvature has the sign variations.

We were unable to find analytical solutions for the system (5a), (5b). Thus, the numerical modeling of this system has been performed. By substituting $\dot{\vec{k}}$ from the first equation (5a) into the second equation (5b), and by substituting $\dot{\vec{r}}$ from the second equation into the first one, we obtain the systems of equations for $k_x$, $k_y$, $x$ and $y$:

$$\begin{cases} \dot{x} = \dfrac{1}{\left(1+\dfrac{\Omega_z(k_x,k_y)}{l_b^2}\right)} \dfrac{1}{\hbar} \dfrac{\partial E_m}{\partial k_x} \\ \dot{y} = \dfrac{1}{\left(1+\dfrac{\Omega_z(k_x,k_y)}{l_b^2}\right)} \dfrac{1}{\hbar} \dfrac{\partial E_m}{\partial k_y} \end{cases} \quad (8a)$$

$$\begin{cases} \dot{k}_x = -\dfrac{1}{\left(1+\dfrac{\Omega_z(k_x,k_y)}{l_b^2}\right)} \dfrac{1}{\hbar l_b^2} \dfrac{\partial E_m}{\partial k_y} \\ \dot{k}_y = \dfrac{1}{\left(1+\dfrac{\Omega_z(k_x,k_y)}{l_b^2}\right)} \dfrac{1}{\hbar l_b^2} \dfrac{\partial E_m}{\partial k_x} \end{cases} \quad (8б)$$

where $l_b = \sqrt{\dfrac{\hbar c}{eB}}$ is the magnetic length. During the integration one needs firstly to obtain the trajectory in $\boldsymbol{k}$-space, and then, by applying the obtained dependencies for $k_x(t)$ and $k_y(t)$, one can find the trajectory in $\boldsymbol{r}$-space. We put $\Delta=0$ for the present integration.

The trajectories in $\boldsymbol{k}$- and $\boldsymbol{r}$-spaces for magnetic field $B$=1MGs are shown in Fig.2(a),(b). It can be seen that the trajectories have the $C_{3V}$ symmetry, and the trajectories in $\boldsymbol{k}$- and $\boldsymbol{r}$-spaces are connected by the $\pi/2$ rotation and by the scale change, as it is in the conventional case.

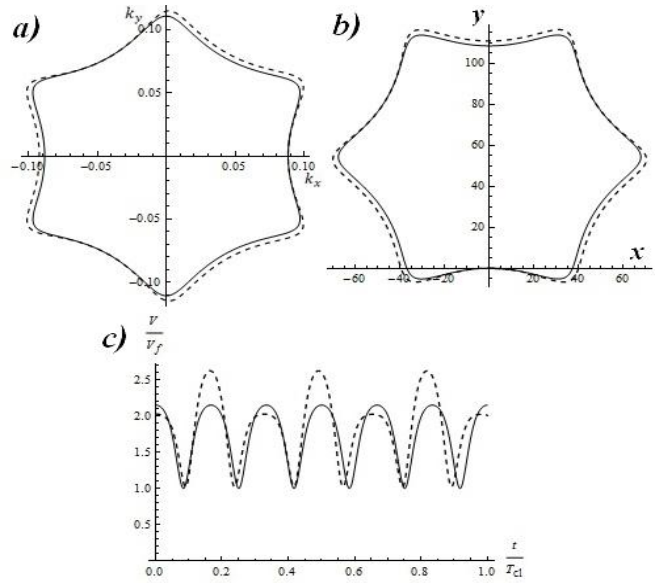

Fig.2. (a) – Electron trajectories in **k**-space, (b) - in **r**-space. (c) – Dependence of the velocity magnitude on time. Bold lines are for the absence of Berry curve and the energy term which is proportional to magnetic moment, and the dashed lines are for the case when these terms are included.

By comparison of the velocities shown in Fig.2 and drawn both with and without the Berry phase term, we can see that the Berry curvature changes the symmetry of the velocity vector field from $C_{6V}$ to $C_{3V}$. It can be noticed also that in the presence of the Berry curvature and the $-\vec{m}\cdot\vec{B}$ term in energy three periods of the velocity oscillations can be observed during one turn, and not six.

One can define analytically the cyclotron frequency in the presence of the Berry curvature for small **k** when the bending effects can be neglected by setting λ=0 in Eq.(6). As a result one has

$$\omega(k) = \frac{1}{1+\frac{\Omega_z(k)}{l_b^2}}\frac{1}{l_b^2}\frac{v}{k}. \tag{8}$$

The dependence of the cyclotron frequency as a function of $k$ is shown in Fig.3. As it follows from the plot and from Eq.(8), near the certain value of $k$ the frequency rapidly increases and changes its sign. The position of this point can be shifted by varying the amplitude of the magnetic field or the energy gap Δ. It can be seen that for large values of $k$ the two curves practically coincide.

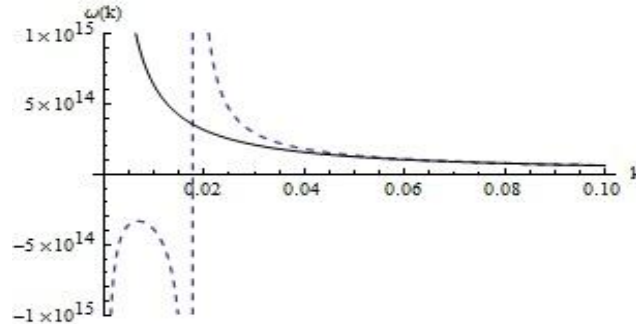

Fig.3. Dependence of the cyclotron frequency on $k$ in the presence of the Berry curvature (dashed line) and in the absence of it (solid line), the parameter $\Delta=0.03эB$.

We may conclude that the interval for $k$ is divided into two regions, where for small $k$ the rotation is in one direction while for big $k$ it is in the opposite direction. The effects considered here can be observed during the investigation of electron cyclotron resonance for surface states in 3D topological insulators.

In case when an energy band is degenerate, the single band approximation is no longer adequate. In this case the wavepacket should be constructed from the Bloch functions from several degenerate bands. The quasiclassical equations of motion for this case have been obtained in [10]. In order for the packet not to split into several parts during its motion it is necessary to construct the wavepacket in the absence of the external fields from the Bloch states from different bands but with the same energy. The equations of dynamics in the magnetic field in such case require special analysis.

To conclude, in the presence of the Berry curvature and the magnetic moment term in energy in the equations of motion for the wavepacket, the symmetry of the electron trajectories is changed, and new resonance cyclotron frequencies appear. Experimentally such changes in the dynamics can be seen during the observation of the cyclotron resonance of surface electrons in 3D topological insulators. Despite the fact that by applying strong magnetic field we destroy the time reversal symmetry, the results of such experiments can lead to the conclusion whether the present isolator was trivial or nontrivial at the zero field.

The authors are grateful to A.A. Perov and D.V. Khomitsky for useful discussions.
The work is supported by the RFBR Grant No. 15-02-04028.

[1] X. L. Qi, S.-C. Zhang, Rev. Mod. Phys. **83**, 1057 (2011).
[2] D. Xiao, M.-C. Chang, Q. Niu, Rev. Mod. Phys. **82,** 1959 (2010).
[3] M.-C. Chang, and Q. Niu, Phys. Rev. B **53,** 7010 (1996).
[4] G.Sundaran, and Q. Niu, Phys. Rev. B **59**, 14915, (1999).
[5] J.N. Fuchs, F. Piechon, M.O. Goerbig, and G Montambaux, Eur. Phys. J. B **77**, 351 (2010)
[6] L. Fu, Phys. Rev. Lett. **103**, 266801 (2009).
[7] Y. Chen, J.G. Analitis, J.-H. Chu, Z. K. Liu, S.-K. Mo, X. L. Qi, H. J. Zhang, D.H. Lu, X. Dai, Z. Fang, S.-C. Zhang, I. R. Fisher, Z. Hussain,and Z.-X. Shen, Science **325**, 178 (2009).
[8] K. Kuroda, M. Arita, K. Miyamoto, M. Ye, J. Jiang, A. Kimura, E.E. Krasovskii, E.V. Chulkov, H. Iwasava, T. Okuda, K. Shimada, Y. Ueda, H. Namatame, and M.Taniguchi, Phys. Rev, Lett. **105**, 076802 (2010).
[9] M. Nomura, S. Souma, A. Takayama, T. Sato, T. Takahashi, K. Eto, K. Segava, and Y. Ando, Phys. Rav. B **89**, 045134 (2014).
[10] D.Culcer, Y.Yao, and Q. Niu Phys.Rev. B, **72**, 085110 (2005).